\documentclass[aps,prd,twocolumn,amsmath,amssymb,showpacs]{revtex4}

\usepackage{graphicx}
\usepackage{dcolumn}
\usepackage{bm}

\begin{document}

\title{FKWC-bases and geometrical identities  \\
for classical and quantum field theories in curved spacetime}

\author{Yves D\'ecanini}
\email{decanini@univ-corse.fr} \affiliation{UMR CNRS 6134 SPE,
Equipe Physique Semi-Classique (et) de la Mati\`ere Condens\'ee, \\
Universit\'e de Corse, Facult\'e des Sciences, BP 52, 20250 Corte,
France}

\author{Antoine Folacci}
\email{folacci@univ-corse.fr} \affiliation{UMR CNRS 6134 SPE,
Equipe Physique Semi-Classique (et) de la Mati\`ere Condens\'ee, \\
Universit\'e de Corse, Facult\'e des Sciences, BP 52, 20250 Corte,
France}

\date{\today}

\begin{abstract}

Fulling, King, Wybourne and Cummings (FKWC) have proposed to expand
systematically the Riemann polynomials encountered in the context of
field theories in curved spacetime on standard bases constructed
from group theoretical considerations. They have also displayed such
bases for scalar Riemann polynomials of order eight or less in the
derivatives of the metric tensor and for tensorial Riemann
polynomials of order six or less. Here we provide a slightly
modified version of the FKWC-bases as well as an important list of
geometrical relations we have used in recent works. These relations
are independent of the dimension of spacetime. In our opinion, they
are helpful to achieve quickly and easily, by hand, very tedious
calculations as well as, of course, to provide irreducible
expressions for all the results obtained. They could be very helpful
to people working in gravitational physics and more particularly (i)
to treat some aspects of the classical theory of gravitational waves
such as the radiation reaction problem and (ii) to deal with
regularization and renormalization of quantum field theories, of
stochastic semiclassical gravity and of higher-order theories of
gravity.

\end{abstract}

\pacs{04.62.+v}

\maketitle

Calculations which must be carried out in the context of field
theories defined on curved spacetimes are non-trivial due to the
systematic occurrence, in the expressions involved, of Riemann
polynomials. These polynomials are formed from the Riemann tensor by
covariant differentiation, multiplication and contraction and their
complexity, degree and number rapidly increase with the precision of
the approximations needed or with the dimension of the gravitational
background considered. Furthermore, the results of these
calculations may be darkened because of the non-uniqueness of their
final forms. Indeed, the symmetries of the Riemann tensor as well as
Bianchi identities can not be used in a uniform manner and monomials
formed from the Riemann tensor may be linearly dependent in
non-trivial ways. In Ref.~\cite{FKWC1992}, Fulling, King, Wybourne
and Cummings (FKWC) have proposed to cure these problems by
expanding systematically the Riemann polynomials encountered in
calculations on standard bases constructed from group theoretical
considerations. They have also displayed such bases for scalar
Riemann polynomials of order eight or less in the derivatives of the
metric tensor and for tensorial Riemann polynomials of order six or
less.

The FKWC-paper has been mainly used in supergravity and string
theory but, unfortunately, it has been rather ignored by people
working in the context of field theories in curved spacetime. This
is a pity because it could be very helpful in order to simplify
considerably a lot of calculations. In the present paper, we shall
provide a slightly modified version of the FKWC-bases as well as a
list of geometrical relations we have used in recent articles
\cite{DecaniniFolacci2005b,DecaniniFolacci2007} dealing with
renormalization of the stress-energy tensor or which, in our
opinion, are important bearing in mind many other aspects of
classical and quantum field theories in curved spacetime. The
results displayed here are independent of the dimension of spacetime
and they have permitted us to achieve rather quickly and easily, by
hand, very tedious calculations as well as to provide irreducible
expressions for all our results. The slight modification of the
FKWC-bases has been mainly done in order to accelerate these
calculations and more particularly the steps involving covariant
differentiation and contraction of Riemann polynomials. We do not
intend to publish our results in a regular scientific journal but we
have decided to distribute them through the arXiv because we are
confident they could be very helpful for other people working in
gravitational physics and more particularly (i) to treat some
aspects of the classical theory of gravitational waves such as the
radiation reaction problem and (ii) to deal with regularization and
renormalization of quantum field theories, of stochastic
semiclassical gravity and of higher-order theories of gravity.

In this paper, we shall use the FKWC-notations ${\cal R}^r_{s,q}$
and ${\cal R}^r_{\lbrace{\lambda_1 \dots \rbrace}}$ to denote
respectively the space of Riemann polynomials of rank r (number of
free indices), order s (number of differentiations of the metric
tensor) and degree q (number of factors $\nabla^p
R_{\dots}^{\dots}$) and the space of Riemann polynomials of rank r
spanned by contractions of products of the type
$\nabla^{\lambda_1}R_{\dots}^{\dots}$. We refer to the FKWC-article
\cite{FKWC1992} for more precisions on these notations and rigor on
the subject. Furthermore, we shall use the geometrical conventions
of Hawking and Ellis \cite{HawkingEllis} concerning the definitions
of the scalar curvature $R$, the Ricci tensor $R_{\mu \nu}$ and the
Riemann tensor $R_{\mu \nu \rho \sigma}$. The geometrical identities
we shall display below and which could be helpful in order to
eliminate ``spurious" Riemann monomials of order up to six by
expressing them in terms of elements of FKWC-bases have been derived
(i) from the commutation of covariant derivatives in the form
\begin{eqnarray}\label{CD_NabNabTensor}
&  & T^{\rho \dots}_{\phantom{\rho} \sigma \dots ;\nu \mu} -
T^{\rho \dots}_{\phantom{\rho} \sigma \dots ;\mu \nu} = \nonumber \\
&   & \qquad  + R^{\rho}_{\phantom{\rho} \tau \mu \nu} T^{\tau
\dots}_{\phantom{\tau} \sigma \dots } + \dots -
R^{\tau}_{\phantom{\tau} \sigma \mu \nu} T^{\rho
\dots}_{\phantom{\rho} \tau \dots} - \dots
\end{eqnarray}
and (ii) from the ``symmetry" properties of the Ricci and the
Riemann tensors (pair symmetry, antisymmetry, cyclic symmetry)
\begin{subequations}
\begin{eqnarray}
& & R_{ab}=R_{ba}  \label{SymRicci} \\
& & R_{abcd}=R_{cdab}  \label{SymRiemann_1} \\
& & R_{abcd}=-R_{bacd} \quad \mathrm{and} \quad R_{abcd}=-R_{abdc}
\label{SymRiemann_2} \\
& & R_{abcd}+R_{adbc}+R_{acdb}=0  \label{SymRiemann_Cycl}
\end{eqnarray}
\end{subequations} as well as (iii) from the Bianchi identity and its consequences
obtained by contraction of index pairs
\begin{subequations}
\begin{eqnarray}
& & R_{abcd;e}+R_{abec;d}+R_{abde;c}=0 \label{AppBianchi_1} \\
& & R_{ abcd}^{\phantom{abcd};d}= R_{ac;b}-R_{bc;a}  \label{AppBianchi_2a} \\
& &R_{ ab}^{\phantom{ab};b}= (1/2)  R_{;a}. \label{AppBianchi_2b}
\end{eqnarray}
\end{subequations}

\subsection{FKWC-basis for Riemann polynomials of rank 0 (scalars) and orders up to six}

The most general expression for a scalar of order six or less in
derivatives of the metric tensor is obtained by expanding it on the
FKWC-basis for Riemann polynomials of rank 0 and order 6 or less.
This basis is composed of the 22 following elements or scalar
Riemann monomials \cite{FKWC1992}:

\noindent - Sub-basis for Riemann polynomials of rank 0 and order 2
(1 element).
\begin{eqnarray} \label{FKWC R0_O2}
&   {\cal R}^0_{2,1}:  &  R
\end{eqnarray}

\noindent - Sub-basis for Riemann polynomials of rank 0 and order 4
(4 elements).
\begin{eqnarray} \label{FKWC R0_O4}
&   {\cal R}^0_{4,1}:  &  \Box R \nonumber \\
&   {\cal R}^0_{4,2}: &  R^2   \quad  R_{pq} R^{pq}  \quad R_{pqrs}
R^{pqrs}
\end{eqnarray}

\noindent - Sub-basis for Riemann polynomials of rank 0 and order 6
(17 elements).
\begin{eqnarray} \label{FKWC R0_O6}
&    {\cal R}^0_{6,1}:  & \Box \Box R     \nonumber \\
& {\cal R}^0_{\lbrace{2,0\rbrace}}:  & R\Box R \quad   R_{;p q}
R^{pq} \quad R_{pq} \Box R^{pq} \quad R_{pq ; rs}R^{prqs} \nonumber \\
&   {\cal R}^0_{\lbrace{1,1 \rbrace}}: & R_{;p}R^{;p} \,\,\,\,
R_{pq;r} R^{pq;r} \,\,\,\, R_{pq;r} R^{pr;q} \,\,\,\, R_{pqrs;t}
R^{pqrs;t}
    \nonumber \\
&    {\cal R}^0_{6,3}:  & R^3   \quad  RR_{pq} R^{pq}  \quad R_{pq}
R^{p}_{\phantom{p} r}R^{qr}  \quad R_{pq}R_{rs}R^{prqs}
 \nonumber \\
& &    RR_{pqrs} R^{pqrs} \quad  R_{pq}R^p_{\phantom{p} rst} R^{qrst
} \nonumber \\
& &  R_{pqrs}R^{pquv} R^{rs}_{\phantom{rs} uv}  \quad R_{prqs} R^{p
\phantom{u} q}_{\phantom{p} u \phantom{q} v} R^{r u s v}
\end{eqnarray}

This basis is a natural one and is often used in this form in
literature. However, it should be noted that certain authors prefer
to use the scalar monomial $R_{pqrs} \Box R^{pqrs}$ instead of the
scalar monomial $R_{pq ; rs}R^{prqs}$. This is the case of Gilkey in
Refs.~\cite{Gilkey75,Gilkey84}. This choice is only a matter of
taste because these two terms appear equally in the calculations
carried out in field theory and the elimination of one of them can
be achieved by using the identity
\begin{eqnarray}\label{Simpl_R0_O6_1}
&  R_{pqrs} \Box R^{pqrs} = 4 \, R_{pq ; rs}R^{prqs}
+2 \, R_{pq}R^p_{\phantom{p} rst} R^{qrst } \nonumber \\
&  \qquad - \, R_{pqrs}R^{pq}_{\phantom{pq} uv} R^{rsuv} - 4\,
R_{prqs} R^{p \phantom{u} q}_{\phantom{p} u \phantom{q} v} R^{r u s
v}
\end{eqnarray}

In the context of field theories in curved spacetime, other Riemann
monomials of rank 0 and orders 4 or 6 such as $R^{pqrs}R_{rqps}$,
$R_{pq}R^{pr}_{\phantom{pr};qr}$, $R^{prqs}R_{pquv}R_{rs
\phantom{uv}}^{\phantom{rs} uv}$, $R_{pq}^{\phantom{pq}
rs}R^{puqv}R_{rusv}$ and $R_{prqs}R^{puqv}R_{\phantom{r} v
\phantom{s} u}^{r \phantom{u} s \phantom{v} }$ can be encountered.
They are not in the FKWC-basis (\ref{FKWC R0_O2})-(\ref{FKWC R0_O6})
and they can be eliminated (i.e., expanded on the FKWC-basis) from
the geometrical identities
\begin{equation}\label{Simpl_R0_O4_1}
R^{pqrs}R_{rqps}=\frac{1}{2} \, R^{pqrs}R_{pqrs}
\end{equation}
\begin{equation}\label{Simpl_R0_O6_2}
R_{pq}R^{pr;q}_{\phantom{pr;q}r} = \frac{1}{2} \, R_{;p q}
R^{pq}+R_{pq} R^{p}_{\phantom{p} r}R^{qr}-R_{pq}R_{rs}R^{prqs}
\end{equation}
and \begin{subequations}
\begin{eqnarray}\label{Simpl_R0_O6_3}
& & R^{prqs}R_{pquv}R_{rs \phantom{uv}}^{\phantom{rs} uv} =
\frac{1}{2}
\, R_{pqrs}R^{pquv} R^{rs}_{\phantom{rs} uv}  \label{Simpl_R0_O6_3a} \\
& & R_{pq}^{\phantom{pq} rs}R^{puqv}R_{rusv} = \frac{1}{4} \,
R_{pqrs}R^{pquv} R^{rs}_{\phantom{rs} uv}  \label{Simpl_R0_O6_3b} \\
& &R_{prqs}R^{puqv}R_{\phantom{r} v \phantom{s} u}^{r \phantom{u} s
\phantom{v} } = -\frac{1}{4} \, R_{pqrs}R^{pquv}
R^{rs}_{\phantom{rs} uv} \nonumber \\
& & \qquad \qquad \qquad + R_{prqs} R^{p \phantom{u} q}_{\phantom{p}
u \phantom{q} v} R^{r u s v}  \label{Simpl_R0_O6_3c}
\end{eqnarray}
\end{subequations}

\subsection{FKWC-basis for Riemann polynomials of rank 1 (vectors) and orders up to 6}

The most general expression for a covariant vector of order six or
less in derivatives of the metric tensor is obtained by expanding it
on the FKWC-basis for Riemann polynomials of rank 1 and order 6 or
less. This basis is composed of the 8 following elements or vector
Riemann monomials \cite{FKWC1992}:

\noindent - Sub-basis for Riemann polynomials of rank 1 and order 3
(1 element).
\begin{eqnarray} \label{FKWC R1_O3}
&   {\cal R}^1_{3,1}:  &  R_{;a}
\end{eqnarray}

\noindent - Sub-basis for Riemann polynomials of rank 1 and order 5
(7 elements).
\begin{eqnarray} \label{FKWC R1_O5}
&   {\cal R}^1_{5,1}:  &  (\Box R)_{;a} \nonumber \\
&   {\cal R}^1_{5,2}: & RR_{;a} \quad R^{;p}R_{p a} \quad R^{pq}
R_{pq ;a} \quad R^{pq} R_{p a ; q}
\nonumber \\
& &  R^{pq;r}R_{rqp a} \quad R^{pqrs} R_{pqrs ;a}
\end{eqnarray}

Here, it should be noted that we have slightly modified the
FKWC-basis of Ref.~\cite{FKWC1992}. Indeed, we have replaced the
term $R^{pq;r} R_{prq a}$ proposed in Ref.~\cite{FKWC1992} by its
opposite given by $R^{pq;r} R_{rqp a}$ for obvious mnemotechnic
reasons.

In the context of field theories in curved spacetime, other Riemann
monomials of rank 1 and order 5 such as $\Box (R_{;a})$, $(\Box
R^p_{\phantom{p} a})_{;p}$ and $R^{pqrs}R_{pqra;s}$ can be
encountered. They are not in the FKWC-basis (\ref{FKWC
R1_O3})-(\ref{FKWC R1_O5}) and they can be eliminated (i.e.,
expanded on the FKWC-basis) from the geometrical identities
\begin{eqnarray}\label{Simpl_R1_O5_1}
& &  \Box (R_{;a}) =  (\Box R)_{;a} + R^{;p}R_{p a}
\end{eqnarray}
and
\begin{eqnarray}\label{Simpl_R1_O5_2}
& &  (\Box R^p_{\phantom{p} a})_{;p} = \frac{1}{2} \, (\Box R)_{;a}
+R^{;p}R_{p a} - R^{pq} R_{pq ;a} \nonumber \\
& & \qquad +2\,  R^{pq} R_{p a ; q} +2\, R^{pq;r}R_{rqp a}
\end{eqnarray}
and
\begin{eqnarray}\label{Simpl_R1_O5_3}
& & R^{pqrs}R_{pqra;s}=\frac{1}{2} \, R^{pqrs}R_{pqrs;a}
\end{eqnarray}

\subsection{FKWC-basis for Riemann polynomials of rank 2 and orders up to 6}

The most general expression for a tensor of rank 2 (more precisely
of type (0,2)) and of order six or less in derivatives of the metric
tensor is obtained by expanding it on the FKWC-basis for Riemann
polynomials of rank 2 and order 6 or less. This basis is composed of
the 29 following elements \cite{FKWC1992}:

\noindent - Sub-basis for Riemann polynomials of rank 2 and order 2
(1 element).
\begin{eqnarray}\label{FKWC R2_O2}
&    {\cal R}^2_{2,1}: &  R_{ab}
\end{eqnarray}

\noindent - Sub-basis for Riemann polynomials of rank 2 and order 4
(6 elements).
\begin{eqnarray}\label{FKWC R2_O4}
&    {\cal R}^2_{4,1}: &  R_{;ab} \quad \Box R_{a b} \nonumber \\
&  {\cal R}^2_{4,2}:  & R R_{a b} \quad R_{p a} R^{p}_{\phantom{p}b}
\quad R^{pq}R_{p a q b} \quad R^{pqr}_{\phantom{pqr} a} R_{pqr b} \nonumber \\
&
\end{eqnarray}

\noindent - Sub-basis for Riemann polynomials of rank 2 and order 6
(22 elements).
\begin{eqnarray}\label{FKWC R2_O6}
&    {\cal R}^2_{6,1}: &  (\Box R)_{;ab} \quad  \Box \Box R_{a b} \nonumber \\
&  {\cal R}^2_{\lbrace{2,0\rbrace}}:  & R R_{;a b} \,\,\,\,  (\Box
R) R_{a b} \,\,\,\,   R_{;p a} R^{p}_{\phantom{p}b} \,\,\,\, R \Box
R_{a b}
\,\,\,\, R_{p a} \Box R^p_{\phantom{p} b} \nonumber \\
& &  R^{pq} R_{pq;a b} \quad
 R^{pq} R_{p a ; bq}  \quad R^{pq} R_{ab; pq} \quad R^{;pq}R_{p
a q b} \nonumber \\
& &   (\Box R^{pq})R_{p a q b} \quad R^{pq;r}_{\phantom{pq;r} a}
R_{rqp b} \quad R^{p \phantom{a}; qr}_{\phantom{p } a} R_{pqr  b}
\nonumber \\
& &  R^{pqrs} R_{pqrs ; a b
} \nonumber \\
&  {\cal R}^2_{\lbrace{1,1 \rbrace}}:  & R_{;a} R_{;b} \quad R_{;p }
R^p_{\phantom{p} a;b} \quad R_{;p } R_{ab}^{\phantom{ab}; p} \quad
R^{pq}_{\phantom{pq};a}R_{pq;b}
   \nonumber \\
& &  R^{pq}_{\phantom{pq} ; a} R_{b p;q}  \quad R^p_{\phantom{p}
a;q} R_{p b}^{\phantom{p b};q} \quad R^p_{\phantom{p} a;q}
R^q_{\phantom{q} b;p}  \nonumber \\
& &  R^{pq;r} R_{rqp a;b } \quad  R^{pq;r} R_{p a q b;r} \quad
R^{pqrs}_{\phantom{pqrs};a} R_{pqrs ; b } \nonumber \\
& & R^{pqr}_{\phantom{pqr}a;s}R_{pqr b}^{\phantom{pqr
b};s}   \nonumber \\
&  {\cal R}^2_{6,3}: & R^2 R_{ab} \quad R R_{p a} R^p_{\phantom{p}b}
\quad R^{pq}R_{pq}R_{ab} \quad R^{pq}R_{p a}R_{q b} \nonumber \\
& &  R R^{pq}R_{p a q b} \quad   R^{pr}R^q_{\phantom{q} r}R_{p a q
b} \quad R^{pq}R^r_{\phantom{r}  a}R_{ rqp  b }
 \nonumber \\
& & R R^{pqr}_{\phantom{pqr}a }R_{pqr b} \quad  R_{ab}R^{pqrs}
R_{pqrs } \quad R^p_{\phantom{p} a}R^{qrs}_{\phantom{qrs} p }R_{ qrs
b } \nonumber \\
& & R^{pq}R^{rs}_{\phantom{rs} pa}R_{rs q b} \quad R_{pq}R^{p r q
s}R_{r a s b} \nonumber \\
& & R_{pq}R^{p rs}_{\phantom{p rs}a}R^q_{\phantom{q} rs b} \quad
R^{pq rs}R_{pq t a }R_{rs \phantom{t} b}^{\phantom{rs} t}
 \nonumber \\
& & R^{p r q s}R^t_{\phantom{t} pq a}R_{t rs b}  \quad
R^{pqr}_{\phantom{pqr} s } R_{pqr t}R^{s \phantom{a} t}_{\phantom{s}
a \phantom{t} b}
\end{eqnarray}

Here again, we have slightly modified the FKWC-basis of
Ref.~\cite{FKWC1992}:

\qquad i) In ${\cal R}^2_{\lbrace{2,0\rbrace}}$ we have replaced the
term $R^{pq} R_{p a ; q b}$ proposed in Ref.~\cite{FKWC1992} by the
term $R^{pq} R_{p a ; b q}$. We think it is more interesting to work
with the latter which directly reduces to an element of the scalar
basis (\ref{FKWC R0_O2})-(\ref{FKWC R0_O6}) by contraction on the
free indices $a$ and $b$. In fact, these two terms are linked by the
geometrical identity
\begin{equation}\label{Simpl_R2_O6_1}
 R^{pq} R_{p a ;q b}=R^{pq} R_{p a ; bq}+R^{pr}R^q_{\phantom{q} r}R_{p a q b}
+R^{pq}R^r_{\phantom{r} a}R_{rqp b}
\end{equation}
so it is easy to return to the original FKWC-basis.

\qquad ii) We have replaced the term $R^{pq;r}_{\phantom{pq;r} a}
R_{prq b}$ of ${\cal R}^2_{\lbrace{2,0\rbrace}}$, the term $R^{pq;r}
R_{prq a;b}$ of ${\cal R}^2_{\lbrace{1,1\rbrace}}$ and the term
$R^{pq}R^r_{\phantom{r} a}R_{prq b}$ of ${\cal R}^2_{6,3}$ proposed
in Ref.~\cite{FKWC1992} by their opposites respectively given by
$R^{pq;r}_{\phantom{pq;r} a} R_{rqp b}$, $R^{pq;r} R_{rqp a;b}$ and
$R^{pq}R^r_{\phantom{r} a}R_{rqp b}$. This choice has been done for
obvious mnemotechnic reasons.

In the context of field theories in curved spacetime, other Riemann
monomials of rank 2 and orders 4 or 6 which are not in the
FKWC-basis (\ref{FKWC R2_O2})-(\ref{FKWC R2_O6}) are systematically
encountered. They can be eliminated (i.e., expanded on the
FKWC-basis) more or less trivially by using more particularly the 15
following geometrical identities:
\begin{eqnarray}\label{Simpl_R2_O4_1}
& R^{pqr}_{\phantom{pqr} a} R_{rqp b} = \frac{1}{2} \,
R^{pqr}_{\phantom{pqr} a} R_{pqr b}
\end{eqnarray} and
\begin{eqnarray}\label{Simpl_R2_O6_2}
& \Box (R_{;ab}) = (\Box R)_{;ab} + 2 \, R_{;p (a}
R^{p}_{\phantom{p}b)} \nonumber \\
& \qquad - 2\, R^{;pq}R_{p a q b} - R_{;p } R_{ab}^{\phantom{ab}; p}
+2 \, R_{;p } R^p_{\phantom{p}(a;b)}
\end{eqnarray}
and
\begin{equation}\label{Simpl_R2_O6_3}
R^{p \phantom{a}; q r}_{\phantom{p } a} R_{r q p b}=\frac{1}{2} \,
R^s_{\phantom{s} a}R^{p q r}_{\phantom{p q r} s}R_{p q r b}-
\frac{1}{2} \, R^{p q}R^{rs}_{\phantom{rs} p a}R_{rs q b}
\end{equation}
and
\begin{eqnarray}\label{Simpl_R2_O6_4}
& & R^{pq;r}R_{raqb;p}= R^{pq;r}R_{rqpb;a} + R^{pq;r}R_{paqb;r}
\end{eqnarray}
and \begin{subequations}
\begin{eqnarray}\label{Simpl_R2_O6_5}
& & R^{pqrs}_{\phantom{pqrs} ;a} R_{pqr b;s}= \frac{1}{2} \,
R^{pqrs}_{\phantom{pqrs} ;a} R_{pqrs;b}, \label{Simpl_R2_O6_5a}\\
& & R^{prs}_{\phantom{prs}a ;q} R^q_{\phantom{q} sr b;p}=
\frac{1}{4} \, R^{pqrs}_{\phantom{pqrs} ;a} R_{pqrs;b}
\label{Simpl_R2_O6_5b}
\end{eqnarray}
\end{subequations}
 and
 \begin{subequations}
\begin{eqnarray}\label{Simpl_R2_O6_6}
&  &  R^{pqrs} R_{pqr a;sb}= \frac{1}{2} \, R^{pqrs} R_{pqrs ; a b}   \label{Simpl_R2_O6_6a} \\
& & R^{pqrs} R_{pqr a;b s}= \frac{1}{2} \, R^{pqrs} R_{pqrs ; a b} -
R^{pqr}_{\phantom{pqr} s } R_{pqr t}R^{s \phantom{a} t}_{\phantom{s}
a \phantom{t} b}  \nonumber \\
& & \qquad  + \frac{1}{2} \, R^{pq rs}R_{pq t a }R_{rs \phantom{t}
b}^{\phantom{rs} t}  +2 \, R^{p r q s}R^t_{\phantom{t} pq a}R_{t rs
b} \label{Simpl_R2_O6_6b}
\end{eqnarray}
\end{subequations}
and
\begin{eqnarray}\label{Simpl_R2_O6_7}
& & R^{pqr}_{\phantom{pqr}a} \Box R_{pqr b}= -2
\,R^{pq;r}_{\phantom{pq;r} b} R_{rqp a}-2 \, R^{p \phantom{b};
qr}_{\phantom{p } b} R_{pqr a} \nonumber \\
& & \qquad + R^p_{\phantom{p} b}R^{qrs}_{\phantom{qrs} p }R_{qrs a}
+R^{pq}R^{rs}_{\phantom{rs} pa}R_{rs q b} \nonumber \\
& & \qquad - R^{pq rs}R_{pq t a }R_{rs \phantom{t} b}^{\phantom{rs}
t} - 4 \,   R^{p r q s}R^t_{\phantom{t} pq a}R_{t rs b}
\end{eqnarray}
and
\begin{eqnarray}\label{Simpl_R2_O6_8}
&  R^{pq}\Box R_{p a q b}= R^{pq} R_{pq;(a b)} -2 \, R^{pq} R_{p (a
;
b)q}  + R^{pq} R_{ab; pq}  \nonumber \\
&  \qquad -2 \, R^{pq}R^r_{\phantom{r} (a}R_{|rqp| b)}  -2 \,
R^{pq}R^{rs}_{\phantom{rs} pa}R_{rs q b}  \nonumber \\
&  \qquad -2 \, R_{pq}R^{p r q s}R_{r a s b} +2 \, R_{pq}R^{p
rs}_{\phantom{p rs}a}R^q_{\phantom{q} rs b}
\end{eqnarray}
\noindent and \begin{subequations}
\begin{eqnarray}\label{Simpl_R2_O6_9}
& & R_{p q}R^{p r s}_{\phantom{p r s}a}R^q_{\phantom{q} s r b} =
R_{p
q}R^{p r s}_{\phantom{p r s}a}R^q_{\phantom{q} r s b} \nonumber \\
& & \qquad \qquad  -\frac{1}{2}
\, R^{p q}R^{r s}_{\phantom{r s} p a}R_{r s q b}  \label{Simpl_R2_O6_9a} \\
& & R_{p q} R^{rsp}_{\phantom{rsp}a}R^q_{\phantom{q} rs b} =
-\frac{1}{2} \, R^{p q}R^{r s}_{\phantom{r s} p a}R_{r s q b}
\label{Simpl_R2_O6_9b}
\end{eqnarray}
\end{subequations}
\noindent and
 \begin{subequations}
\begin{eqnarray} \label{Simpl_R2_O6_10}
& & R^{p q r s}R^t_{\phantom{t} p q a }R_{t r s b} = \frac{1}{4} \,
R^{p q r s}R_{p q t a }R_{r s \phantom{t} b}^{\phantom{r s} t}
\label{Simpl_R2_O6_10a} \\
& & R^{p q r  s}R_{p q t a }R^t_{\phantom{t} r s b} = - \frac{1}{2}
\, R^{p q r s}R_{p q t a }R_{r s \phantom{t} b}^{\phantom{r s} t}
\label{Simpl_R2_O6_10b} \\
& & R^{p r q s}R_{p q t a }R_{r s \phantom{t} b}^{\phantom{r s} t} =
\frac{1}{2} \, R^{p q r s}R_{p q t a }R_{r s \phantom{t}
b}^{\phantom{r s} t}  \label{Simpl_R2_O6_10c} \\
& & R^{p r q s}R^t_{\phantom{t} p q a }R_{t s r b} = R^{p r q
s}R^t_{\phantom{t} p q a }R_{t r s b} \nonumber \\
& & \qquad \qquad - \frac{1}{4} \, R^{p q r s}R_{p q t a }R_{r s
\phantom{t} b}^{\phantom{r s} t}
\label{Simpl_R2_O6_10d} \\
& & R^{p r q  s}R_{p q t a }R^t_{\phantom{t} s r b} = \frac{1}{4} \,
R^{p q r s}R_{p q t a }R_{r s \phantom{t} b}^{\phantom{r s} t}
\label{Simpl_R2_O6_10e}
\end{eqnarray}
\end{subequations}
It should be noted that there also exists a lot of other relations
involving terms cubic in the Riemann tensor which are useful in
calculations but they can be obtained trivially from the five
previous ones.

\subsection{FKWC-basis for Riemann polynomials of rank 3 and orders up to 6}

The most general expression for a tensor of rank 3 (more precisely
of type (0,3)) and of order six or less in derivatives of the metric
tensor is obtained by expanding it on the FKWC-basis for Riemann
polynomials of rank 3 and order 6 or less. This basis is composed of
the 12 following elements \cite{FKWC1992}:

\noindent - Sub-basis for Riemann polynomials of rank 3 and order 3
(1 element).
\begin{eqnarray}\label{FKWC R3_O3}
&    {\cal R}^3_{3,1}: &  R_{ab;c}
\end{eqnarray}

\noindent - Sub-basis for Riemann polynomials of rank 3 and order 5
(11 elements).
\begin{eqnarray}\label{FKWC R3_O5}
&    {\cal R}^3_{5,1}: &  R_{;abc} \quad (\Box R_{a b})_{;c} \nonumber \\
&  {\cal R}^3_{5,2}:  & R_{;a} R_{bc} \quad RR_{ab;c} \quad
R^p_{\phantom{p} a} R_{pb;c} \quad R^p_{\phantom{p} a} R_{bc;p}
\nonumber \\
& &  R^{;p}R_{p a b c} \quad R^{pq}_{\phantom{pq} ;a} R_{p bqc}
\quad R^{p \phantom{a} ;q}_{\phantom{p} a}  R_{pb qc} \nonumber \\
& & R^{pq}R_{paqb;c} \quad R^{pqr}_{\phantom{pqr} a} R_{pqr b;c}
\end{eqnarray}

Here, we have modified the FKWC-basis of Ref.~\cite{FKWC1992} by
choosing to omit their $R^{p \phantom{a} ;q}_{\phantom{p} a} R_{pq
bc}$ term which is explicitly antisymmetric on $b$ and $c$ and which
vanishes by contraction on these indices because it can be expanded
as
\begin{equation}\label{Simpl R3_O5 0a}
R^{p \phantom{a} ;q}_{\phantom{p} a} R_{pq bc}=R^{p \phantom{a}
;q}_{\phantom{p} a}  R_{pb qc} - R^{p \phantom{a} ;q}_{\phantom{p}
a}  R_{pc qb}.
\end{equation}
In Ref.~\cite{FKWC1992}, Fulling, King, Wybourne and Cummings have
kept such a term because they have chosen to consider that $R^{p
\phantom{a} ;q}_{\phantom{p} a} R_{pb qc}$ does not generate any
contribution antisymmetric in $b$ and $c$.

In the context of field theories in curved spacetime, other Riemann
monomials of rank 3 and order 5 such as $\Box (R_{ab;c})$,
$R^{pqr}_{\phantom{pqr} a}R_{rbqc;p}$ and $R^{pqr}_{\phantom{pqr}
a}R_{rqpb;c}$ can be encountered. They are not in the FKWC-basis
(\ref{FKWC R3_O3})-(\ref{FKWC R3_O5}) and they can be eliminated
(i.e., expanded on the FKWC-basis) from the geometrical identities

\begin{eqnarray}\label{Simpl_R3_O5_1}
&  \Box (R_{ab;c}) =  (\Box R_{a b})_{;c} + 2\, R^p_{\phantom{p} (a}
R_{|pc|;b)}  - 2 \, R^p_{\phantom{p} (a} R_{b)c;p} \nonumber \\
& \quad + R^p_{\phantom{p} c} R_{ab;p} -4\, R^{p \phantom{(a}
;q}_{\phantom{p} (a}  R_{|p|b) qc}
\end{eqnarray}
\noindent and \begin{subequations}
\begin{eqnarray}\label{Simpl_R3_O5_2}
&  R^{pqr}_{\phantom{pqr} a}R_{rbqc;p} = - \frac{1}{2} \,
R^{pqr}_{\phantom{pqr} a} R_{pqrb;c}  \label{Simpl_R3_O5_2a} \\
& R^{pqr}_{\phantom{pqr} a}R_{rqpb;c} =\frac{1}{2} \,
R^{pqr}_{\phantom{pqr} a} R_{pqrb;c}  \label{Simpl_R3_O5_2b}
\end{eqnarray}
\end{subequations}

\subsection{FKWC-basis for Riemann polynomials of rank 4 and orders up to 6}

The most general expression for a tensor of rank 4 (more precisely
of type (0,4)) and of order six or less in derivatives of the metric
tensor is obtained by expanding it on the FKWC-basis for Riemann
polynomials of rank 4 and order 6 or less. This basis is composed of
the 48 following elements \cite{FKWC1992}:

\noindent - Sub-basis for Riemann polynomials of rank 4 and order 2
(1 element).
\begin{eqnarray}\label{FKWC R4_O2}
&    {\cal R}^4_{2,1}: &  R_{abcd}
\end{eqnarray}

\noindent - Sub-basis for Riemann polynomials of rank 4 and order 4
(5 elements).
\begin{eqnarray}\label{FKWC R4_O4}
&    {\cal R}^4_{4,1}: &  R_{ab;cd} \nonumber \\
&    {\cal R}^4_{4,2}: &  R_{ab}R_{cd} \quad RR_{abcd} \quad
R^p_{\phantom{p} a}R_{pbcd} \quad  R^{p \phantom{a} q}_{\phantom{p}
a \phantom{q} b}R_{pcqd} \nonumber \\
&
\end{eqnarray}

\noindent - Sub-basis for Riemann polynomials of rank 4 and order 6
(42 elements).
\begin{eqnarray}\label{FKWC R4_O6}
&    {\cal R}^4_{6,1}: &  R_{;abcd} \quad (\Box R_{ab})_{;cd}    \nonumber \\
&    {\cal R}^4_{\lbrace{2,0\rbrace}}: & R_{;ab}R_{cd} \quad
RR_{ab;cd} \quad R_{ab} \Box R_{cd} \quad R^p_{\phantom{p}
a}R_{pb;cd} \nonumber \\
& & R^p_{\phantom{p} a}R_{bc;dp} \quad (\Box R)R_{abcd} \quad
R^{;p}_{\phantom{;p} a}R_{pbcd}  \nonumber \\
& & (\Box R^{p}_{\phantom{p} a})R_{pbcd} \quad R^{pq}_{\phantom{pq}
;ab}R_{pcqd} \quad  R^{p \phantom{a;b} q}_{\phantom{p}a;b}R_{pcqd}
\nonumber \\
& & R_{ab}^{\phantom{ab};pq}R_{pcqd} \quad R^{pq}R_{paqb;cd} \quad
R^{pqr}_{\phantom{pqr}
a}R_{pqrb;cd} \nonumber \\
&    {\cal R}^4_{\lbrace{1,1\rbrace}}: & R_{;a} R_{bc;d} \quad
R^p_{\phantom{p} a;b} R_{pc;d} \quad R^p_{\phantom{p} a;b} R_{cd;p}
\quad   R_{ab}^{\phantom{ab};p}R_{cd;p} \nonumber \\
& & R^{;p}R_{pabc;d} \quad R^{pq}_{\phantom{pq} ;a} R_{p bqc;d}
\quad R^{p \phantom{a} ;q}_{\phantom{p} a}  R_{pbqc;d}  \nonumber \\
& &  R^{pqr}_{\phantom{pqr} a;b} R_{pqr c;d} \quad
R^{p \phantom{a} q \phantom{b} ;r}_{\phantom{p} a \phantom{q} b}R_{pcqd;r}\nonumber \\
&    {\cal R}^4_{6,3}: & RR_{ab}R_{cd} \quad
R_{ab}R^p_{\phantom{p}c}R_{pd} \quad R^2R_{abcd} \quad
RR^p_{\phantom{p}a}R_{pbcd} \nonumber \\
& & R^{pq}R_{pq}R_{abcd} \quad R_{ab}R^{pq}R_{pcqd} \quad
R^p_{\phantom{p}a}R^q_{\phantom{q}b}R_{pcqd} \nonumber \\
& & R^{pq}R_{pa}R_{qbcd} \quad  RR^{p \phantom{a} q}_{\phantom{p} a
\phantom{q} b}R_{pcqd}
\quad R_{ab}R^{pqr}_{\phantom{pqr}c}R_{pqrd} \nonumber \\
& & R^p_{\phantom{p}a}R^{r\phantom{p} s}_{\phantom{r} p \phantom{s}
 b}R_{rcsd}  \quad  R^{pq}R^r_{\phantom{r} pqa}R_{rbcd} \nonumber \\
 & & R^{pq}R^r_{\phantom{r} apb}R_{r cqd} \quad R_{abcd}R^{pqrs}R_{pqrs} \nonumber \\
 & &
R^{qrs}_{\phantom{qrs}p}R^p_{\phantom{p}abc}R_{qrs d}
 \quad R^{prqs}R_{paqb}R_{rcsd} \nonumber \\
& &  R^{p \phantom{a} q}_{\phantom{p}a \phantom{q}
b}R^{rs}_{\phantom{rs}pc}R_{rs qd} \quad R^{p \phantom{a}
q}_{\phantom{p}a \phantom{q} b}R^{\phantom{p} rs}_{p \phantom{rs}
c}R_{q rs d}
\end{eqnarray}

It should be noted that we have modified the FKWC-basis of
Ref.~\cite{FKWC1992}:

\qquad i) In ${\cal R}^4_{\lbrace{2,0\rbrace}}$ we have replaced the
term $R^p_{\phantom{p} a}R_{bc;pd}$ proposed in Ref.~\cite{FKWC1992}
by the term $R^p_{\phantom{p} a}R_{bc;dp}$ because it is more
interesting to work with the latter which directly reduces to an
element of the basis (\ref{FKWC R2_O2})-(\ref{FKWC R2_O6}) by
contraction on the free indices $c$ and $D$. In fact, it is easy to
return to the original FKWC-basis by using the geometrical identity
\begin{equation}\label{Simpl_R4_O6_1}
 R^p_{\phantom{p} a}R_{bc;pd}=R^p_{\phantom{p}
a}R_{bc;dp}+ R^p_{\phantom{p}a}R^q_{\phantom{q}b}R_{pdqc} +
R^p_{\phantom{p}a}R^q_{\phantom{q}c}R_{pdqb}
\end{equation}
which links these two Riemann monomials.

\qquad ii) In ${\cal R}^4_{\lbrace{2,0\rbrace}}$ we have replaced
the term $R^{p \phantom{a};q }_{\phantom{p}a \phantom{;q}
b}R_{pcqd}$ proposed in Ref.~\cite{FKWC1992} by the term $R^{p
\phantom{a;b} q}_{\phantom{p}a;b}R_{pcqd}$ which directly reduces to
an element of the basis (\ref{FKWC R2_O2})-(\ref{FKWC R2_O6}) by
contraction on the free indices $a$ and $b$. It is easy to return to
the original FKWC-basis by using the geometrical identity
\begin{eqnarray}\label{Simpl_R4_O6_2}
& & R^{p \phantom{a};q }_{\phantom{p}a \phantom{;q} b}R_{pcqd}= R^{p
\phantom{a;b} q}_{\phantom{p}a;b}R_{pcqd} -
R^p_{\phantom{p}a}R^{r\phantom{p} s}_{\phantom{r} p \phantom{s}
 b}R_{rcsd} \nonumber \\
 & & \qquad \qquad \qquad + R^{pq}R^r_{\phantom{r} bpa}R_{r dqc}
\end{eqnarray}
which links these two Riemann monomials.

\qquad iii) We have furthermore replaced the terms $R^{pq}R_{p
\phantom{r} qa}^{\phantom{p} r} R_{rbcd}$ and $R^{pq}R_{p
\phantom{r} ab}^{\phantom{p} r}R_{qcrd}$ of ${\cal R}^4_{6,3}$ which
have been proposed in Ref.~\cite{FKWC1992} by their opposites
$R^{pq}R^r_{\phantom{r} pqa}R_{rbcd}$, $R^{pq}R^r_{\phantom{r}
pab}R_{r cq d}$ for obvious mnemotechnic reasons.

\qquad iv) We have finally omitted in the original FKWC-basis of
Ref.~\cite{FKWC1992} the term $R^{pq}_{\phantom{pq} ab}R_{pqcd}$ in
${\cal R}^4_{4,2}$, the term $R^{p \phantom{a} ;q}_{\phantom{p} a}
R_{pq bc;d}$ in ${\cal R}^4_{\lbrace{1,1\rbrace}}$ and the terms
$R^p_{\phantom{p}a}R^q_{\phantom{q}b}R_{pqcd}$,
$RR^{pq}_{\phantom{pq}ab}R_{pq cd}$,
$R^p_{\phantom{p}a}R^{rs}_{\phantom{rs} pb}R_{rscd}$,
$R^{pq}R^r_{\phantom{r} pab}R_{r qcd}$, $R^{pq}R^r_{\phantom{r}
pab}R_{r cq d}$, $R^{pqrs}R_{pqab}R_{rscd}$,
$R^{pq}_{\phantom{pq}ab}R^{rs}_{\phantom{rs}pc}R_{rs qd}$ and
$R^{pq}_{\phantom{pq}ab}R^{\phantom{p} rs}_{p \phantom{rs} c}R_{q rs
d}$ in ${\cal R}^4_{6,3}$ because we prefer to expand them on the
remaining terms as
\begin{equation}\label{Simpl R4_O4 0a}
R^{pq}_{\phantom{pq} ab}R_{pqcd} = 2\, R^{p \phantom{a}
q}_{\phantom{p} a \phantom{q} b}R_{pcqd}- 2\, R^{p \phantom{a}
q}_{\phantom{p} a \phantom{q} b}R_{pdqc}
\end{equation}
and
\begin{equation}\label{Simpl R4_O6 0a}
R^{p \phantom{a} ;q}_{\phantom{p} a} R_{pq bc;d} = R^{p \phantom{a}
;q}_{\phantom{p} a}  R_{pbqc;d}-R^{p \phantom{a} ;q}_{\phantom{p} a}
R_{pcqb;d}
\end{equation}
and
\begin{equation}\label{Simpl R4_O6 0b}
R^p_{\phantom{p}a}R^q_{\phantom{q}b}R_{pqcd} =
R^p_{\phantom{p}a}R^q_{\phantom{q}b}R_{pcqd}-R^p_{\phantom{p}a}R^q_{\phantom{q}b}R_{pdqc}
\end{equation}
and
\begin{equation}\label{Simpl R4_O6 0c}
 RR^{pq}_{\phantom{pq}ab}R_{pq cd} = 2\, RR^{p \phantom{a}
q}_{\phantom{p} a \phantom{q} b}R_{pcqd}- 2\, RR^{p \phantom{a}
q}_{\phantom{p} a \phantom{q} b}R_{pdqc}
\end{equation}
and
\begin{equation}\label{Simpl R4_O6 0d}
R^p_{\phantom{p}a}R^{rs}_{\phantom{rs} pb}R_{rscd} = 2\,
R^p_{\phantom{p}a}R^{r\phantom{p} s}_{\phantom{r} p \phantom{s}
 b}R_{rcsd}- 2\, R^p_{\phantom{p}a}R^{r\phantom{p} s}_{\phantom{r} p \phantom{s}
 b}R_{rdsc}
\end{equation}
and \begin{subequations}
\begin{eqnarray}\label{Simpl R4_O6 0e}
&R^{pq}R^r_{\phantom{r} pab}R_{r qcd}= R^{pq}R^r_{\phantom{r}
apb}R_{r cqd} - R^{pq}R^r_{\phantom{r} bpa}R_{r cqd} \nonumber \\
& \quad
- R^{pq}R^r_{\phantom{r} apb}R_{r dqc} + R^{pq}R^r_{\phantom{r} bpa}R_{r dqc} \label{Simpl R4_O6 0e1}\\
& R^{pq}R^r_{\phantom{r} pab}R_{r cq d} =  R^{pq}R^r_{\phantom{r}
apb}R_{r cqd}- R^{pq}R^r_{\phantom{r} bpa}R_{r cqd} \nonumber
\\
& \label{Simpl R4_O6 0e2}
\end{eqnarray}
\end{subequations}
and
\begin{equation}\label{Simpl R4_O6 0f}
R^{pqrs}R_{pqab}R_{rscd}= 4\, R^{prqs}R_{paqb}R_{rcsd} - 4\,
R^{prqs}R_{paqb}R_{rdsc}
\end{equation}
and \begin{subequations}
\begin{eqnarray}\label{Simpl R4_O6 0g}
& & R^{pq}_{\phantom{pq}ab}R^{rs}_{\phantom{rs}pc}R_{rs qd} = R^{p
\phantom{a} q}_{\phantom{p}a \phantom{q}
b}R^{rs}_{\phantom{rs}pc}R_{rs qd} - R^{p \phantom{a}
q}_{\phantom{p}a \phantom{q} b}R^{rs}_{\phantom{rs}pd}R_{rs
qc}\nonumber
\\
& & \label{Simpl R4_O6 0g1}\\
& & R^{pq}_{\phantom{pq}ab}R^{\phantom{p} rs}_{p \phantom{rs} c}R_{q
rs d} = R^{p \phantom{a} q}_{\phantom{p}a \phantom{q}
b}R^{\phantom{p} rs}_{p \phantom{rs} c}R_{q rs d} - R^{p \phantom{a}
q}_{\phantom{p}a \phantom{q} b}R^{\phantom{p} rs}_{p \phantom{rs}
d}R_{q rs c}   \nonumber
\\
& & \label{Simpl R4_O6 0g2}
\end{eqnarray}
\end{subequations}

In the context of field theories in curved spacetime, other Riemann
monomials of rank 4 and orders 4 or 6 which are not in the
FKWC-basis (\ref{FKWC R4_O2})-(\ref{FKWC R4_O6}) are systematically
encountered. They can be eliminated (i.e., expanded on the
FKWC-basis) more or less trivially by using the ``symmetry"
properties of the Ricci and the Riemann tensors
(\ref{SymRicci})-(\ref{SymRiemann_Cycl}), the Bianchi identity and
its consequences (\ref{AppBianchi_1})-(\ref{AppBianchi_2b}) as well
as from the commutation of covariant derivatives in the form
(\ref{CD_NabNabTensor}). Among the multitude of geometrical
identities which can be obtained in this context we have retained
more particularly the following ones:
\begin{eqnarray}\label{Simpl_R4_O4_NEW1}
&  R^{p \phantom{a} q}_{\phantom{p} a \phantom{q} b}R_{pqcd} = R^{p
\phantom{a} q}_{\phantom{p} a \phantom{q} b}R_{pcqd} -  R^{p
\phantom{a} q}_{\phantom{p} a \phantom{q} b}R_{pdqc}
\end{eqnarray}
and
\begin{eqnarray}\label{Simpl_R4_O6_NEW1}
& &  \Box R_{abcd} =  R_{ac;bd} - R_{bc;ad} -R_{ad;bc} + R_{bd;ac} \nonumber \\
& & \qquad + R^p_{\phantom{p} c}R_{pdab} - R^p_{\phantom{p}
d}R_{pcab}\nonumber \\
& & \qquad +2\, R^{p \phantom{a} q}_{\phantom{p} a \phantom{q}
d}R_{pbqc} - 2\, R^{p \phantom{a} q}_{\phantom{p} a \phantom{q}
c}R_{pbqd} \nonumber \\
& & \qquad - 2\, R^{p \phantom{a} q}_{\phantom{p} a \phantom{q}
b}R_{pcqd} + 2\, R^{p \phantom{a} q}_{\phantom{p} a \phantom{q}
b}R_{pdqc}
\end{eqnarray}
and \begin{subequations}
\begin{eqnarray}\label{Simpl_R4_O6_3}
&  R^{pqr}_{\phantom{pqr} a;b}R_{rqpc;d} =  \frac{1}{2} \,
R^{pqr}_{\phantom{pqr} a;b}R_{pqr c;d}  \label{Simpl_R4_O6_3a} \\
& R^{pqr}_{\phantom{pqr} a;b}R_{rcqd;p} =- \frac{1}{2} \,
R^{pqr}_{\phantom{pqr} a;b} R_{pqrc;d}  \label{Simpl_R4_O6_3b}
\end{eqnarray}
\end{subequations}
\noindent and \begin{subequations}
\begin{eqnarray}\label{Simpl_R4_O6_4}
&  R^{pqr}_{\phantom{pqr} a}R_{rqpb;cd} =  \frac{1}{2} \,
R^{pqr}_{\phantom{pqr} a}R_{pqr b;cd}  \label{Simpl_R4_O6_4a} \\
& R^{pqr}_{\phantom{pqr} a}R_{rbqc;pd} =- \frac{1}{2} \,
R^{pqr}_{\phantom{pqr} a}R_{pqr b;cd}  \label{Simpl_R4_O6_4b}
\end{eqnarray}
\end{subequations}
\noindent and
\begin{eqnarray}\label{Simpl_R4_O6_5}
& & R^{p \phantom{a} q}_{\phantom{p} a \phantom{q} b} \Box R_{pcqd}=
R^{pq}_{\phantom{pq} ;cd}R_{paqb} -  R^{p \phantom{c;d}
q}_{\phantom{p}c;d}R_{pbqa}
\nonumber \\
& & \quad   -  R^{p \phantom{d;c} q}_{\phantom{p}d;c}R_{paqb} +
R_{cd}^{\phantom{cd};pq}R_{paqb}
\nonumber \\
& & \quad +  R^p_{\phantom{p}c}R^{r\phantom{p} s}_{\phantom{r} p
\phantom{s}
 d}R_{rbsa}  + R^p_{\phantom{p}d}R^{r\phantom{p} s}_{\phantom{r} p
\phantom{s}
 c}R_{rbsa}  \nonumber \\
& & \quad + R^{pq}R^r_{\phantom{r} apb}R_{r cqd}-
R^{pq}R^r_{\phantom{r} apb}R_{r dqc} \nonumber \\
& & \quad  - 2\, R^{prqs}R_{paqb}R_{rcsd}
\nonumber \\
& & \quad  - R^{p \phantom{a} q}_{\phantom{p}a \phantom{q}
b}R^{rs}_{\phantom{rs}pc}R_{rs qd} - R^{p \phantom{a}
q}_{\phantom{p}a \phantom{q} b}R^{rs}_{\phantom{rs}pd}R_{rs qc} \nonumber \\
& & \quad +2\, R^{p \phantom{a} q}_{\phantom{p}a \phantom{q}
b}R^{\phantom{p} rs}_{p \phantom{rs} d}R_{q rs c}
\end{eqnarray}
and
\begin{equation}\label{Simpl_R4_O6_NEW2}
R^p_{\phantom{p}a}R^{rs}_{\phantom{rs} pb}R_{rcsd} =
R^p_{\phantom{p}a}R^{r\phantom{p} s}_{\phantom{r} p \phantom{s}
 b}R_{rcsd}-   R^p_{\phantom{p}a}R^{r\phantom{p} s}_{\phantom{r} p \phantom{s}
 b}R_{rdsc}
\end{equation}
 and
 \begin{subequations}
\begin{eqnarray} \label{Simpl_R4_O6_6}
& & R^{p \phantom{a} q}_{\phantom{p}a \phantom{q}
b}R^{rs}_{\phantom{rs}pc}R_{q rs d} = -\frac{1}{2} \, R^{p
\phantom{a} q}_{\phantom{p}a \phantom{q}
b}R^{rs}_{\phantom{rs}pc}R_{rs qd}
\label{Simpl_R4_O6_6a} \\
& & R^{p \phantom{a} q}_{\phantom{p}a \phantom{q} b}R^{\phantom{p}
rs}_{p \phantom{rs} c}R_{q sr d} = R^{p \phantom{a} q}_{\phantom{p}a
\phantom{q} b}R^{\phantom{p} rs}_{p \phantom{rs} c}R_{q rs d}
\nonumber \\
& & \qquad \qquad - \frac{1}{2} \, R^{p \phantom{a} q}_{\phantom{p}a
\phantom{q} b}R^{rs}_{\phantom{rs}pc}R_{rs qd}
\label{Simpl_R4_O6_6b}
\end{eqnarray}
\end{subequations}
and
\begin{equation}\label{Simpl R4_O6 7}
R^{pqrs}R_{paqb}R_{rcsd}=   R^{prqs}R_{paqb}R_{rcsd} -
R^{prqs}R_{paqb}R_{rdsc}
\end{equation}

\subsection{FKWC-basis for Riemann polynomials of rank 5 and orders up to 6}

The most general expression for a tensor of rank 5 (more precisely
of type (0,5)) and of order six or less in derivatives of the metric
tensor is obtained by expanding it on the FKWC-basis for Riemann
polynomials of rank 5 and order 6 or less. This basis is composed of
the 9 following elements \cite{FKWC1992}:

\noindent - Sub-basis for Riemann polynomials of rank 5 and order 3
(1 element).
\begin{eqnarray}\label{FKWC R5_O3}
&    {\cal R}^5_{3,1}: &  R_{abcd;e}
\end{eqnarray}

\noindent - Sub-basis for Riemann polynomials of rank 5 and order 5
(8 elements).
\begin{eqnarray}\label{FKWC R5_O5}
&    {\cal R}^5_{5,1}: &  R_{ab;cde} \nonumber \\
&    {\cal R}^5_{5,2}: &  R_{ab}R_{cd;e} \quad R_{;a}R_{bcde} \quad
RR_{abcd;e} \quad R^p_{\phantom{p} a;b}R_{pcde} \nonumber
\\
& & R_{ab}^{\phantom{ab} ;p}R_{pcde}    \quad R^p_{\phantom{p}
a}R_{pbcd;e}  \quad R^{p \phantom{a} q}_{\phantom{p} a \phantom{q}
b}R_{pcqd;e}
\end{eqnarray}

Here, we have modified the FKWC-basis of Ref.~\cite{FKWC1992} by
choosing to omit their term $R^{p \phantom{a} ;q}_{\phantom{p} a}
R_{pq bc}$ explicitly antisymmetric in $b$ and $c$ and which
vanishes by contraction on these indices because it can be expanded
as
\begin{equation}\label{Simpl R5_O5 0a}
R^{pq}_{\phantom{pq} ab}R_{pqcd;e}=2\, R^{p \phantom{a}
q}_{\phantom{p} a \phantom{q} b}R_{pcqd;e} - 2\, R^{p \phantom{a}
q}_{\phantom{p} a \phantom{q} b}R_{pdqc;e}.
\end{equation}

Among the multitude of geometrical identities which can be obtained
for the Riemann monomials of rank 5 and order 5 we only retain
\begin{equation}\label{Simpl R5_O5 1}
R^{p \phantom{a} q}_{\phantom{p} a \phantom{q} b} R_{pqcd;e}= R^{p
\phantom{a} q}_{\phantom{p} a \phantom{q} b}R_{pcqd;e} -  R^{p
\phantom{a} q}_{\phantom{p} a \phantom{q} b}R_{pdqc;e}.
\end{equation}

Furthermore, it should be noted that in field theory calculations at
two-loop order in a four dimensional gravitational background or at
one-loop order in higher dimensional background, only three of these
Riemann monomials of rank 5 are used ($R_{ab;cde}$, $R_{ab}R_{cd;e}$
and $R^{p \phantom{a} q}_{\phantom{p} a \phantom{q} b}R_{pcqd;e}$)
because there is no other Riemann monomial of rank 5 present in the
covariant Taylor series expansion of the Van Vleck-Morette
determinant $\Delta^{1/2}$ (see Ref.~\cite{DecaniniFolacci2005a}).

\subsection{FKWC-basis for Riemann polynomials of rank 6 and orders up to 6}

The most general expression for a tensor of rank 6 (more precisely
of type (0,6)) and of order six or less in derivatives of the metric
tensor is obtained by expanding it on the FKWC-basis for Riemann
polynomials of rank 6 and order 6 or less. This basis is composed of
37 elements: 1 element in ${\cal R}^6_{4,1}$, 2 elements in ${\cal
R}^6_{4,2}$, 1 element in ${\cal R}^6_{6,1}$, 9 elements in ${\cal
R}^6_{\lbrace{2,0\rbrace}}$, 6 elements in ${\cal
R}^6_{\lbrace{1,1\rbrace}}$ and 18 elements in ${\cal R}^6_{6,3}$.
This basis is fully described in Ref.~\cite{FKWC1992}. In field
theory calculations at two-loop order in a four dimensional
gravitational background or at one-loop order in higher dimensional
background), we only need the 8 elements of this basis which are
present in the covariant Taylor series expansion of the Van
Vleck-Morette determinant $\Delta^{1/2}$ namely (see
Ref.~\cite{DecaniniFolacci2005a}):

\begin{eqnarray}\label{FKWC R6_O6}
&    \mathrm{in} \, \, {\cal R}^6_{6,1}: &  R_{ab;cdef} \nonumber \\
&    \mathrm{in} \, \, {\cal R}^6_{\lbrace{2,0\rbrace}}: &
R_{ab}R_{cd;ef} \quad  R^{p \phantom{a} q}_{\phantom{p} a
\phantom{q} b}R_{pcqd;ef}  \nonumber \\
&    \mathrm{in} \, \, {\cal R}^6_{\lbrace{1,1\rbrace}}: &
R_{ab;c}R_{de;f} \quad  R^{p \phantom{a} q}_{\phantom{p} a
\phantom{q} b;c}R_{pdqe;f}  \nonumber \\
&    \mathrm{in} \, \, {\cal R}^6_{6,3}: & R_{ab}R_{cd}R_{ef} \quad
R_{ab}R^{p \phantom{c} q}_{\phantom{p} c \phantom{q} d}R_{peqf}
\nonumber \\
& &  R^{p}_{\phantom{p} a q b}R^{q}_{\phantom{q}
crd}R^{r}_{\phantom{r} epf}
\end{eqnarray}

\begin{acknowledgments}
We would like to thank Rosalind Fiamma for help with the English.
\end{acknowledgments}


\bibliography{FKWC-bases_FTinCS}

\end{document}